# Infrared dielectric metamaterials from high refractive index chalcogenides

H. N. S. Krishnamoorthy,[1,†,*] G. Adamo,[1,†] J. Yin,[1] V. Savinov,[2] N. I. Zheludev[1,2], C. Soci[1,*]

1. Centre for Disruptive Photonic Technologies, TPI, SPMS, Nanyang Technological University, Singapore 637371
2. Optoelectronics Research Centre & Centre for Photonic Metamaterials, University of Southampton, SO17 1BJ, UK

[†]Equally contributing authors
*Correspondence: harish.k@ntu.edu.sg, csoci@ntu.edu.sg

**High-index dielectric materials are in great demand for nanophotonic devices and applications, from ultrathin optical elements to metal-free sub-diffraction light confinement and waveguiding. Here we show that chalcogenide topological insulators are particularly apt candidates for dielectric nanophotonics architectures in the infrared spectral range by reporting metamaterial resonances in chalcogenide crystals sustained well inside the mid-infrared, choosing $Bi_2Te_3$ as case study within this family of materials. Strong resonant modulation of the incident electromagnetic field is achieved thanks to the exceptionally high refractive index ranging between 7 and 8 throughout the 2-10 μm region. Analysis of the complex mode structure in the metamaterial allude to the excitation of poloidal surface currents which could open pathways for enhanced light-matter interaction and low-loss plasmonic configurations by coupling to the spin-polarized topological surface carriers, thereby providing new opportunities to combine dielectric, plasmonic and magnetic metamaterials in a single platform.**

Topological insulator (TI) crystals feature time reversal symmetry-protected, highly conducting surface states characterized by Dirac dispersion and spin-momentum locking of carriers that encapsulate a semi-conducting bulk.[1–3] They are an extremely attractive class of materials for electronic[4], spintronic[5] and, more recently, photonic[6,7] applications, where coupling of light to the topologically protected surface carriers may lead to propagating surface plasmon polaritons with very little scattering and other exotic phenomena.[8] This has motivated extensive studies of electromagnetic properties of TI chalcogenide crystals over a broad range of frequencies from THz to UV.[9–15] Interaction of electromagnetic waves with the topological surface states can be enhanced by suitably structuring the TI crystals with subwavelength units, such as resonant metamolecules, giving rise to absorption and localization of the electromagnetic field. TI metamaterials have been realized at THz and UV-visible frequencies,[9,16–18] however there have been hardly any studies on resonant TI structures at intermediate near- and mid-infrared frequencies, where the compositionally tunable refractive index is extremely high,[19] and optical conductivity from charge carriers in topological surface states becomes significant.[20] Within the family of chalcogenide crystals, we select $Bi_2Te_3$ to demonstrate dielectric metamaterial structures in the technologically important near to mid-infrared frequency window. $Bi_2Te_3$ has a refractive index between 7 and 8 over the 2-10 μm spectral range, which is much larger than typical infrared dielectric materials such as Si, Ge and PbTe. We exploit this exceptionally high refractive index to prove resonant behaviour in dielectric nanoslit metamaterials, with distinct modes sustained deep into the mid-infrared region. Analysis of the mode structure suggests the existence of complex higher-order modes within the nanoslits. We argue that engineering of poloidal surface currents associated with these unconventional structural modes opens up new opportunities to couple light with spin-polarized topological surface state carriers.

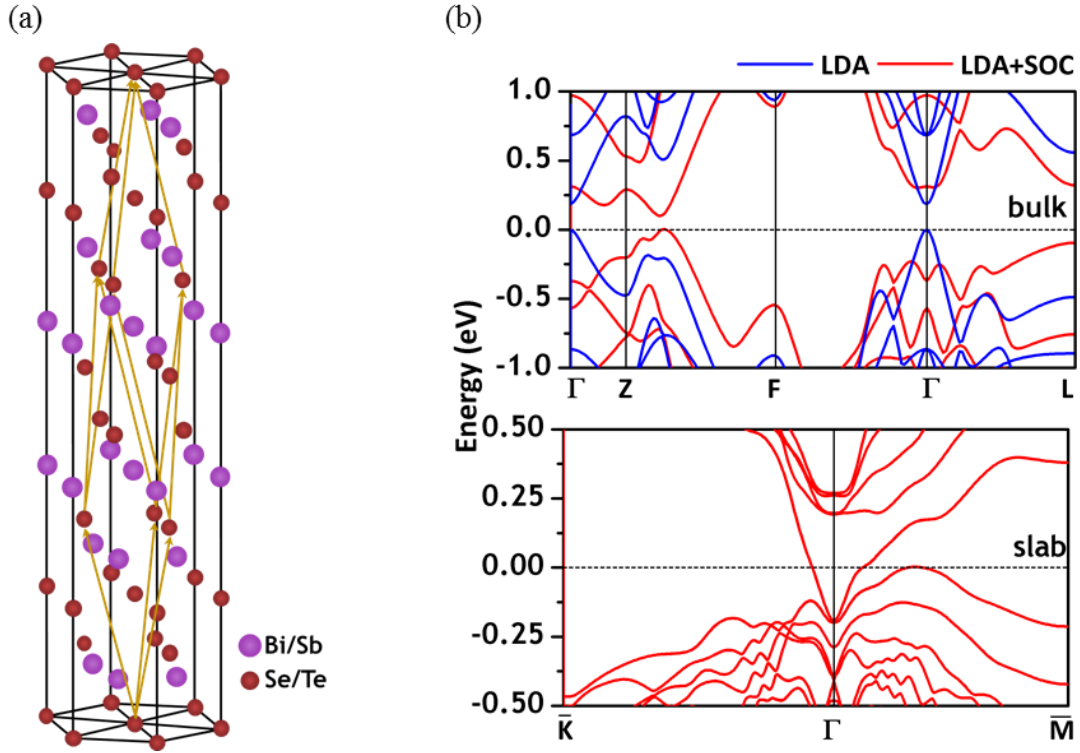

**Figure 1.** (a) Crystal structure of rhombohedral-phase $Bi_2Te_3$. (b) Band structure of $Bi_2Te_3$ bulk (top panel) and slab (bottom panel) calculated at the LDA level, with (red lines) and without (blue lines) spin-orbit coupling. Including the latter leads to band-inversion at the Γ point and gives rise to topologically protected states at the surface (bottom panel).

We started by carrying out first-principles calculations and spectroscopic measurements of single crystal $Bi_2Te_3$. Density functional theory (DFT) calculations based on the local density approximation (LDA) were employed to study the electronic band structures and optical properties of the rhombohedral-phase $Bi_2Te_3$ (see the crystal structure in **Figure 1a**) using the Quantum ESPRESSO (QE) package.[21] Experimental lattice parameters of bulk $Bi_2Te_3$[22,23] were used to build the initial structure, and ground states geometry of the $Bi_2Te_3$ was obtained by the total energy minimization method upon relaxing their crystal framework and atomic coordinates. We calculated the band structure of $Bi_2Te_3$ with and without spin-orbit coupling (SOC). Due to the presence of heavy elements such as Bi and Te, relativistic effects and SOC have significant impact on the band structure. Without SOC, the bands have the typical parabolic dispersion with a direct gap at the Γ point. However, the presence of SOC leads to band-inversion at the Γ point and a topologically non-trivial gap is induced (**Figure 1b top panel**) with the Dirac dispersion appearing in the case of a thin slab (**Figure 1b bottom panel**). The optical response was calculated by employing the Bethe-Salpeter equations (BSE) method with the YAMBO code, using ground-state wavefunctions from QE package.[24,25] The imaginary part of the permittivity was determined by evaluating direct electronic transitions between occupied and higher-energy unoccupied electronic states and the real part was obtained by employing Kramers-Kroning transformation on the imaginary part. Additional details on optical response calculations can be found in reference [20].

**Figure 2a** shows the real and imaginary parts of the calculated complex refractive index ($\tilde{n} = n + ik$). Its dispersion is characterized by contrasting behaviours in two regions of the spectrum: (i) the short wavelength (0.25-1.50 μm) region with strong absorption (high $k$), resulting from interband transitions in the bulk, and (ii) the long wavelength (~2-16 μm) region featuring strong polarizability (high $n$). Correspondingly, the permittivity is negative throughout the 0.25-0.85 μm region ($n < k$),[20,26] where the material is plasmonic. Note that the contribution from surface state carriers to the optical conductivity becomes significant in the long wavelength region above 6 μm, as manifested by a small decrease of the refractive index.

Spectroscopic studies were carried out on exfoliated films of topological insulator single crystal samples of $Bi_2Te_3$ with thickness ranging from 10 to 100 μm. We characterized optical properties over a broad spectral range, from the UV to the mid-infrared by means of variable angle ellipsometry (in the UV to near-infrared range) and near- to mid-infrared reflection measurements, from which we extracted the experimental complex refractive index dispersion (**Figure 2b**). The experimental dispersion captures the essential features derived in the calculated optical constants, showing absorption and negative permittivity at the shorter wavelengths, and strong dielectric behaviour in the infrared.

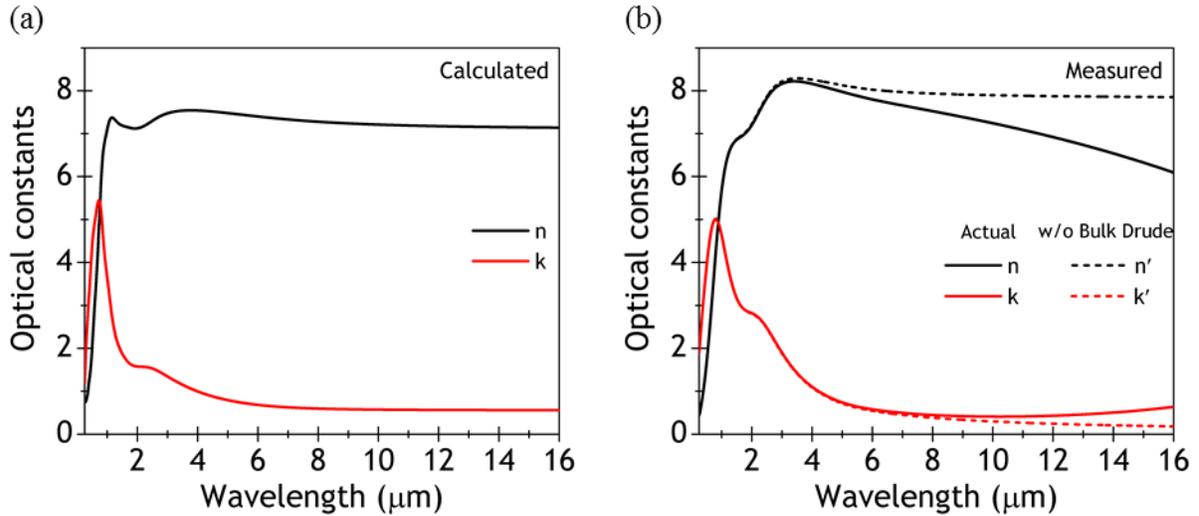

**Figure 2.** Complex refractive index dispersion of the $Bi_2Te_3$ topological insulator crystals used in this study, (a) calculated from first-principles technique and, (b) measured experimentally (solid lines) from ellipsometry and infrared spectroscopy measurements. The dashed lines correspond to the case if the Drude contribution from the free carriers in the valence band is removed.

The deviation from the calculated dispersion at longer wavelengths is due to a sharp decrease of the refractive index and an increase of the extinction coefficient induced by free bulk carriers from intrinsic doping. This effectively creates a third region of high refractive index and low losses between 7 and 10 μm for the crystals in our hands. The dashed lines in Figure 2b show the experimental optical constants after removing the bulk Drude contribution for comparison with the DFT calculation results, showing fairly good agreement between the two once the intrinsic doping contribution is excluded.

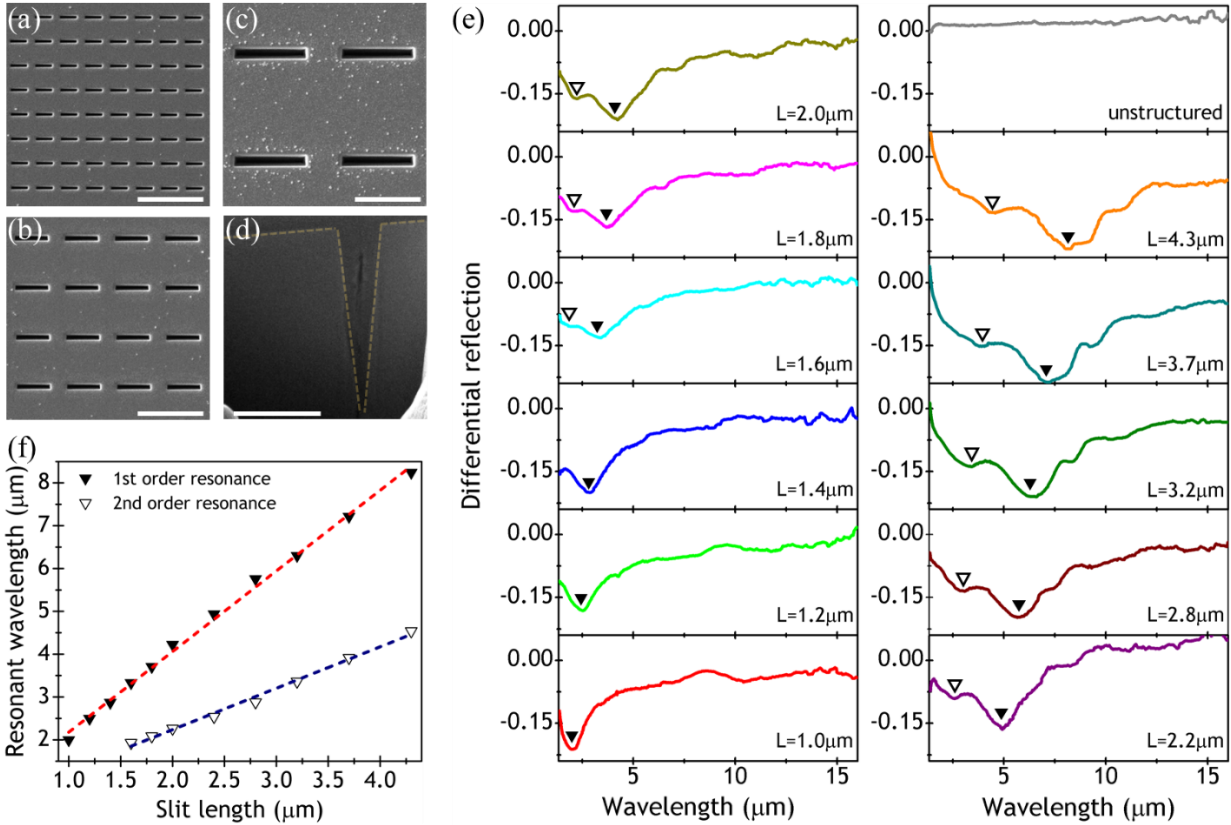

**Figure 3**. (a), (b), and (c) SEM images of slit arrays with lengths L = 1.0 μm, 2.0 μm, and 4.3 μm, respectively. Scale bar corresponds to 4 μm. (d) Cross-sectional image of slit array of length 4.3 μm. Scale bar corresponds to 2 μm. Dashed light yellow lines is an aid to mark the v-shaped contour of the slit. (e) Experimental infrared differential reflection spectra from slit arrays of different slit lengths from 1.0 um to 4.3 um. Pronounced resonances, both first (▼) and second order (▽) are observed that red-shift with the slit length. (f) Resonant wavelength as a function of slit length for the fundamental and second order resonant modes in the nanoslit arrays.

Optical materials with high refractive index and low losses such as $Bi_2Te_3$ in the mid-infrared are in great demand for dielectric metamaterials, as they can produce strong mode confinement and narrow resonances for small form factor devices. On this premise, we fabricated infrared nanoslit arrays via focused ion beam milling on the surface of exfoliated $Bi_2Te_3$ crystals. The slit array geometry was chosen for geometrical simplicity and designed to have pronounced resonances across the entire infrared spectrum by varying the nanoslit length (L) from 1.0 to 4.3 µm. Representative top-view and cross-sectional SEM images of the fabricated nanoslit arrays are shown in **Figures 3a-d**. Infrared microscopy reflection measurements were carried out with incident electric field vector polarized both parallel ($R_{E\parallel}$) and perpendicular ($R_{E\perp}$) to the length of the slits. The slit arrays show resonant response only when excited with perpendicular polarization. To highlight these resonances, we plot the differential reflection spectra, $(R_{E\perp} - R_{E\parallel})/R_{E\parallel}$, for the various slit lengths (**Figure 3e**). Two distinct resonances are clearly observed in the plot, the fundamental resonances (indicated by ▼) and, the second order resonances (indicated by ▽) which appear in the measured spectral region for slits longer than 1.5 µm. **Figure 3f** shows the variation of the resonant wavelength as a function of slit length, clearly indicating a linear spectral red-shift with scaling factors of 2 and 1 times the slit length for the fundamental and second order resonances, respectively.[27]

To further our understanding of the nature of the main resonant mode, we carried out finite element simulations of the nanoslit array, over the corresponding wavelength interval. **Figure 4a** compares the experimental and simulated differential reflection spectra of the longest slit (L=4.3 µm) arrays, showing good agreement between the two. Corresponding maps of the resonant electric and magnetic fields at 8.55 µm are plotted in **Figures 4b** and **4c** for incident electric field polarization perpendicular to the slits. The field-maps for the minor mode at 4.35 µm are plotted in **Figure 4d,e**. Based on field-maps alone one would categorize the mode at 8.55 µm (Figure 4 b,c) as toroidal dipole, due to a vortex of oscillating magnetic field (Figure 4c), however the exact multipole decomposition, shown in **Figure 4f**, shows that the situation is far more nuanced. The metamaterial excitation is not dominated by a particular mode, but is a combination of multiple modes such as the electric dipole (order 0), magnetic dipole and electric quadrupole (order 1), toroidal dipole, magnetic dipole and electric quadrupole (order 2) etc, depicted schematically in **Figure 4g**. The differential reflection spectrum calculated from multipole analysis is shown as dashed lines in Figure 4a and agrees fairly well with the simulated and experimental spectra. For more details of the multipole analysis, refer to Supplementary Information section I. Such high-order modes supported by the metamaterial[28–30] enable strong light localization and confinement, which may be suitable for applications in nonlinear and laser optics,[31,32] or in coupling to high-order transitions in atoms or molecules. In the context of TI metamaterials, the unique nature of the resonant fields featuring poloidal currents on the surface of the TI may be exploited to couple light with spin-polarized carriers and gain optical access to the topological surface states. This becomes particularly relevant in the mid-IR region where chalcogenide crystals feature a combination of high refractive index and larger contribution of topological surface states to the optical conductivity.

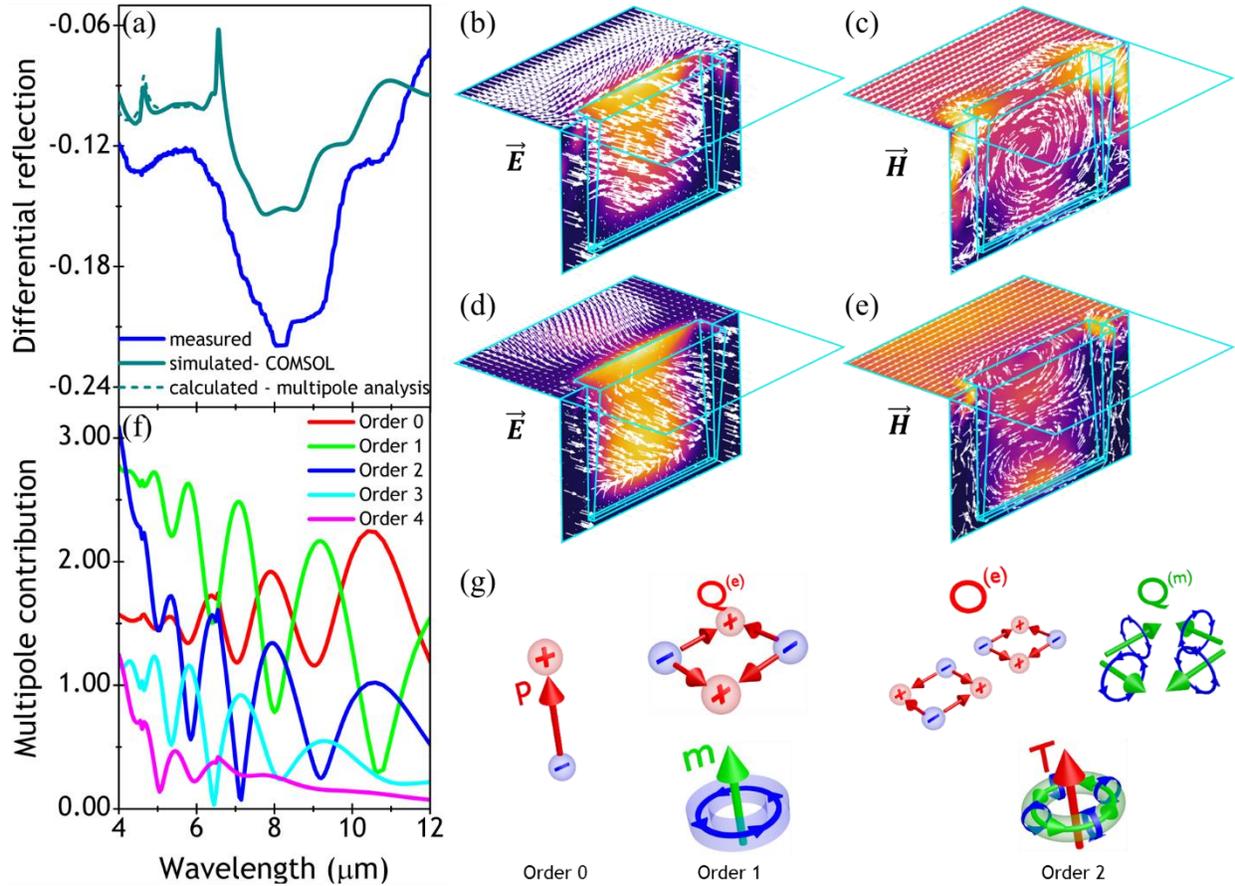

**Figure 4**. (a) Experimental (solid blue), simulated (solid dark cyan) and calculated (dashed lines) differential reflection spectra of $Bi_2Te_3$ metamaterial slit array of length 4.30 μm. The fundamental and second order resonances at 8.55 μm and 4.35 μm, respectively are evident in the plot. The sharp peaks in the simulated and calculated spectra around 4.60 μm and 6.50 μm are due to diffraction effects. Maps of electric ($\vec{E}$) and, magnetic ($\vec{H}$) fields determined by FEM simulations showing the nature of the mode at the, (b), (c) fundamental, and (d), (e) second order resonances. The maps have been calculated over one half of the slit, for simulation size purposes. The cyan outline overlaid on the colour maps indicates the full slit and simulation region for clarity. (f) The contribution of various multipoles to the (displacement) current excitation induced in the metamaterial by the incident wave polarized perpendicular to the slits, including electric dipole (order 0), magnetic dipole (order 1), electric quadrupole (order 1), and toroidal dipole (order 2). (g) Schematic representation of the various modes in the metamaterial. See Supplementary Information for further details.

Overall, the TI chalcogenide crystal family is an exceptionally versatile material platform for infrared applications based on high-index, low-loss dielectric metamaterial architectures, including ultrathin flat optical elements[33], sub-diffraction light confinement and waveguiding[34], and nonlinear optics[35]. Low-loss mid-IR metamaterials are also highly sought for enhanced sensing of molecular fingerprints based on strong light confinement.[36,37] Chalcogenide crystal metamaterials add broadband tunability of the resonances by compositional[38] and structural design to the high refractive index. Moreover, as shown here for $Bi_2Te_3$, the inverse geometry of nanoslits carved in high-index crystals induces higher-order complex modes outside the dielectric medium, which provide additional pathways to sense changes in the surrounding environment.

In conclusion, chalcogenide topological insulator crystals are a compelling materials platform for photonic applications in the infrared part of the spectrum. We have shown that $Bi_2Te_3$ exhibits a strong polarizability with refractive index that exceeds 7 in the 2 – 10 μm range, larger than conventional dielectric materials that aids strong nanostructure resonances sustained deep into the mid-infrared. The exceptionally high index facilitates formation of poloidal currents at the surface of the material that may potentially be coupled to spin-polarized topological surface states. This opens the path to new infrared metamaterials combining dielectric, plasmonic and magnetic properties for applications including molecular fingerprinting, environmental sensing, and integrated mid-IR photonics.

**Methods:**

**Experimental:** $Bi_2Te_3$ crystals were purchased commercially from 2D Semiconductors Inc. The infrared reflection/transmission spectra of the unstructured TIs crystals and microscopy reflection spectra of structured $Bi_2Te_3$ were measured using a Bruker Hyperion microscope coupled to a Bruker Vertex 80v spectrometer. Spectroscopic ellipsometry data were collected using a J. A. Woollam VASE ellipsometer in the 250 nm – 1650 nm spectral range over three angles of incidence (30°, 50° and 70°), and analyzed using the CompleteEASE ellipsometry data analysis program. The refractive index values in the infrared were determined using the experimentally measured near-to-mid-infrared reflection spectrum and the dielectric constants in the UV-near infrared range measured from ellipsometry. This analysis was carried out using the RefFIT program wherein a combination of Tauc-Lorentz and Lorentz oscillators were used to model simultaneously, the ellipsometric dielectric constants as well as the infrared reflection spectrum, from which the refractive index of the material in the infrared spectral range was determined.

**Simulations:** The optical response of the nanoslit array was simulated using full-wave Maxwell equations solver COMSOL. The simulations were carried out for a 3D structure using perfect electric/magnetic conductor boundary conditions on the mirror symmetry plane containing the long axis of the slit for electric field polarized perpendicular/parallel to the slit. In the multipole analysis, the number of higher order modes obtained is dependent on the position of the air-metamaterial interface relative to the z=0 plane. The optimal position of this interface is chosen such that the differential reflection as well as the field induced in the metamaterial can be quantified with the minimum possible number of modes, and corresponds to z=-1200 nm for the mode contributions shown in Figure 4f.

**Acknowledgments**

This research was supported by the Singapore Ministry of Education (Grant Number: MOE2016-T3-1-006), the Agency for Science, Technology and Research (Grant Number: A*STAR-SERC A18A7b0058), and the Engineering and Physical Sciences Research Council, UK (Grant Numbers: EP/M009122/1).